# CIRCA: comprehensible online system in support of chest X-rays-based COVID-19 diagnosis

Wojciech Prazuch[1,#], Aleksandra Suwalska[1,#], Marek Socha[1,#], Joanna Tobiasz[1,2], Pawel Foszner[1,2], Jerzy Jaroszewicz[3], Katarzyna Gruszczynska[3], Magdalena Sliwinska[4], Jerzy Walecki[5], Tadeusz Popiela[6], Grzegorz Przybylski[7], Andrzej Cieszanowski[8], Mateusz Nowak[9], Malgorzata Pawlowska[10], Robert Flisiak[11], Krzysztof Simon[12], Gabriela Zapolska[13], Barbara Gizycka[14], Edyta Szurowska[15], POLCOVID Study Group, Michal Marczyk[1,16], Joanna Polanska[1,*]

[1] Department of Data Science and Engineering, Silesian University of Technology, Gliwice, Poland

[2] Department of Computer Graphics, Vision and Digital Systems, Silesian University of Technology, Gliwice, Poland

[3] Department of Infectious Diseases and Hepatology, Medical University of Silesia, Katowice, Poland

[4] Department of Diagnostic Imaging, Voivodship Specialist Hospital, Wroclaw, Poland

[5] Department of Diagnostic Radiology, Central Clinical Hospital of the Ministry of Internal Affairs and Administration, Warsaw, Poland

[6] Department of Radiology, Jagiellonian University Medical College, Krakow, Poland

[7] Department of Lung Diseases, Cancer and Tuberculosis, Kujawsko-Pomorskie Pulmonology Center, Bydgoszcz, Poland

[8] Department of Imaging Diagnostics, National Institute of Oncology, Warsaw, Poland

[9] Department of Radiology, Silesian Hospital, Cieszyn, Poland

[10] Department of Infectious Diseases and Hepatology, Collegium Medicum in Bydgoszcz, Nicolaus Copernicus University, Torun, Poland

[11] Department of Infectious Diseases and Hepatology, Medical University of Bialystok, Bialystok, Poland

[12] Department of Infectious Diseases and Hepatology, Wroclaw Medical University, Wroclaw, Poland

[13] Department of Radiology, Czerniakowski Hospital, Warsaw, Poland

[14] PDepartment of Imaging Diagnostics, MEGREZ Hospital, Tychy, Poland

[15] 2nd Department of Radiology, Medical University of Gdansk, Poland

[16] Yale Cancer Center, Yale School of Medicine, New Haven, CT, USA

[#] Equal contribution

[*] Corresponding author

# Abstract

**Background:** Due to the large accumulation of patients requiring hospitalization, the COVID-19 pandemic disease caused a high overload of health systems, even in developed countries. Deep learning techniques based on medical imaging data can help in the faster detection of COVID-19 cases and monitoring of disease progression. However, regardless of the numerous proposed solutions using lung X-rays, none of them is a product that can be used in the clinic.

**Methods:** Five different datasets (clinical POLCOVID and AIforCOVID databases, a public collection of images called COVIDx set, pneumonia patients from NIH, and artificially generated data) were used to construct a representative dataset of 23 799 CXRs for model training, 1 050 images as a hold-out test set and 44 247 CXRs as independent test set (BIMCV database). A U-Net-based model was developed to identify a clinically relevant region of the CXR that contains the lungs and each class (normal, pneumonia, and COVID-19) was divided into 3 subtypes using a 2D Gaussian mixture model. A decision tree was used to aggregate predictions from the InceptionV3 convolutional network based on processed CXRs and a dense 7-layer neural network on radiomic features determined independently for the upper, middle, and lower lung fragments.

**Results:** The lung segmentation model gave the Sorensen-Dice coefficient of 94.86% in the validation dataset, and 93.36% in the testing dataset. In 5-fold cross-validation, the accuracy for all classes ranged from 91% to 93%, keeping slightly higher specificity than sensitivity. Also, NPV was higher than PPV for each class to guarantee a low rate of false negative cases. In the hold-out test set, the balanced accuracy ranged between 68% and 100%. The highest performance was obtained for the subtypes N1, P1, and C1, while images assigned to subtypes N3, P3, or C3 have been wrongly predicted among themselves. A similar performance was obtained on the independent dataset for normal and COVID-19 class subtypes, but not for pneumonia class subtypes. Seventy-six percent of COVID-19 patients wrongly classified as normal cases were annotated by radiologists as 'Negative for Pneumonia'.

**Conclusions:** The CIRCA system showed a great performance to distinguish COVID-19 patients from healthy individuals and other pneumonia cases during model training, initial validation, and independent test set. Examination of class subtypes revealed the high clinical heterogeneity of the CXR data, which explained the variability of the results. Finally, we developed and maintain the online service (https://circa.aei.polsl.pl) to provide easy and publicly available access to fast diagnosis support tools.

# Introduction

The first cases of viral pneumonia caused by a novel Severe Acute Respiratory Syndrome Coronavirus 2 (SARS-CoV-2) were reported in the Chinese city of Wuhan at the end of 2019. Since then, this new respiratory tract disease called COVID-19 has spread globally and has caused numerous mortalities worldwide. The present criteria for the diagnosis of COVID-19 include clinical symptoms, epidemiological history, and laboratory testing; however, the final confirmation of this disease is based on the positive result of the reverse-transcription-polymerase-chain-reaction (RT-PCR) test[1]. Although the specificity of this test is very high, the sensitivity is only moderate leading to multiple false negatives, especially at an early stage of the disease. Also, in some regions of the world, the accessibility to a test is limited.

Medical imaging modalities, such as chest X-ray (CXR) and computed tomography (CT), might be an alternative or could complement RT-PCR tests to provide correct and rapid diagnosis since lung abnormalities can be detected very early, even in asymptomatic patients[2,3]. The use of CT in patients with SARS-CoV-2 infection showed high sensitivity[4] but has multiple limitations, e.g. transit of contagious patients to the radiology department, exposure of radiology staff to SARS-CoV-2, contamination of CT scanner, so it is primarily recommended to be used only in specific clinical situations that will change patient management decisions[5]. Chest radiography is easily accessible and may be done portably on a patient bed by a single radiographer. However, the main drawback of this method is lower sensitivity and specificity for the detection of lung lesions early in the disease by human experts, so efficient computer-based solutions are needed. Unfortunately, limited data, especially in COVID-19 cases, made the task of creating a good prognostic model difficult[6].

From the start of the world pandemic in 2020 various deep learning techniques, mostly based on convolutional neural networks (CNN), were applied to detect the disease on CXRs with very high accuracy[7-10]. The proposed structures were highly complex, with a very large number of layers inadequate to the size of the datasets (only hundreds of COVID-19 cases were then provided). The classifiers were also trained to distinguish COVID-19 from normal lungs without considering other pneumonia that might show similar symptoms. When new datasets started to appear, it was found that the models worked very poorly on data not seen during model training (lack of model generalization) which eliminates them from the potential clinical use[11-13]. Multiple reasons for such behavior have been given, but the major ones are: (i) low quality, small size, and reduced clinical diversity of training datasets; (ii) no evaluation of the model on an independent dataset; (iii) inefficient pre-processing of CXRs leading to the dependency of the model on confounding factors rather than medical pathology. For example, some studies used pediatric data to represent normal and pneumonia classes so the models learned probably only age-related features to detect COVID-19[13]. When explainable AI tools were used to understand the models better, like saliency maps, existing solutions incorrectly highlighted regions outside the lung area to be important for COVID-19 classification[12].

Using the recommendations provided in the discussed publications, we propose a complete system with an online open-access platform called CIRCA to distinguish SARS-CoV-2 infection from other types of pneumonia and normal cases on chest X-Ray images. With limited access to COVID-19 CXRs at the stage of network training, we required the lung segmentation to be done independently, as the initial step in data analysis. For classification model building, we used diverse clinical datasets with images gathered at many hospitals from different world regions. The final classification was made using the aggregation of results of two deep learning models: (i) image-based; (ii) radiomics features-based. Finally, the model was validated on the independent clinical dataset, providing additional information about disease subtypes along with a patient diagnosis.

## Materials and Methods

### Super-resolution model development data

High-resolution radiographs gathered from the POLCOVID dataset were utilized for creating a supervised dataset of low- and high-resolution image pairs. POLCOVID dataset was created from radiographs originating from 15 different medical facilities and this made it a good fit for creating a diverse data source for the super-resolution problem. 2 912 images of high-resolution were gathered including 953 COVID-19, 923 pneumonia, and 1 036 normal images. Each radiograph was downsampled to the 512x512 resolution as a standard resolution for web scraped radiographs. Additionally, to be able to analyze very low-quality images, the images were also downsampled to 256x256 resolution.

### Lung segmentation model development data

A dataset consisting of 2 294 publicly available CXRs was gathered. Among them, 1 124 images were collected from the National Institutes of Health – Clinical Center database[14], 662 images from Shenzhen No.3 Hospital in Shenzhen, China[15], 138 from the tuberculosis control program of the Department of Health and Human Services of Montgomery County, USA[15], and 370 images were acquired from a public database of Guangzhou Women and Children's Medical Center, Guangzhou, China[16]. Lung mask annotation for each X-ray image was generated manually by the experts. The left atrium and left ventricle together with the aortic knob were excluded from the mask to ensure proper segmentation of lung tissue.

### Classification model development data

The dataset used in the study was created with images from five different sources: our internal POLCOVID database, COVIDx database[17], AIforCovid database[18], NIH pneumonia database[14], and artificial images generated with Generative Adversarial Network (GAN). Each image represented a chest X-Ray of a patient with one of the three diagnoses: healthy (normal), pneumonia (not related to COVID-19), or COVID-19.

POLCOVID database contained 4 809 images from 15 different medical facilities with 2 426 healthy, 1 147 pneumonia, and 1 236 COVID-19 cases. COVIDx database contained 15 403 images with 8 066 normal, 5 573 pneumonia 1 764 COVID-19 cases. AIforCovid included 1 105 COVID-19 images from six Italian hospitals. NIH provided 2 061 pneumonia cases while 2 635 images resembling COVID-19 cases were generated with GAN.

GAN architecture was composed of two models: a generator that generates new samples and a discriminator that tries to determine if a given sample is real or fake (generated). The generator had one fully connected (dense) and four transposed convolutional layers with the ReLU activation function. After each convolutional layer, a batch normalization with a momentum of 0.8 was used. Images were generated from latent vectors of 256 features. The discriminator was built of four convolutional layers and one fully connected output layer. All but the last layer's activation function in the decoder was leaky ReLU with the alpha parameter of 0.2. The last layer's activation function was sigmoid. The GAN model was trained for 1500 epochs.

### Independent test set

BIMCV COVID-19+[19] and PadChest COVID-19-[20] CXRs datasets served for the independent testing. Both datasets are available on the Valencian Region Medical ImageBank website. Only Posterior-Anterior (PA) and Anterior-Posterior (AP) projections were considered. COVID-19- cases with comorbidities other than pneumonia were removed from the data set. Finally, the test set consisted of 35 857 healthy cases, 2 119 pneumonia cases, and 6 271 COVID-19 cases. The two datasets were connected and termed BIMCV in the analysis.

### Data pre-processing

The raw CXR image undergoes a series of image processing procedures to properly prepare it for the key steps of analysis: lung area segmentation and classification. In the first stage, the images were standardized by analyzing the pixel intensity distribution, where outliers (determined based on calculated 0.0025 and 0.9975 percentiles as cut-off points) were removed and the pixel values were transformed to the range of [0, 1]. These steps allow for uniformity of X-Rays obtained from different hospitals and the removal of some artifacts (e.g., white letters). After the standardization stage, each image was subjected to the contrast enhancement procedure to better expose the border of the lung area of patients with advanced lesions.

### Data cleaning

The data used in the classification task were cleaned by removing: (i) too small images; (ii) poor segmentation images. The classification model uses images after lung segmentation, so the image was considered too small and discarded if at least one of the lung region dimensions was less than 300 pixels. The method for finding images with poor segmentation quality was performed with the following steps. First, four measures quantifying the properties of the binary mask of the lung region were calculated: (i) eccentricity - the ratio of the distance between the focal points of the fitted ellipse to the length of its main axis (range from 0 to 1; higher better); (ii) orientation - the angle between the x-axis and the larger axis of the ellipse (range from -90 to 90; closer to 0 better); (iii) area - the proportion of pixels in the lung region in relation to the entire image (range from 0 to 1; higher better); (iv) solidity - the proportion of pixels in a convex hull, which are also in the region (0 to 1; higher better). The lung segmentation quality score was calculated as the arithmetic mean of these 4 properties and normalized to range from 0 to 1. The threshold for removing poor-quality images was found using the outlier detection method for skewed distributions[21].

### UMAP-based data visualization

To visualize relations between the data a Uniform Manifold Approximation and Projection (UMAP)[22] embedding was created using features extracted from the neural network[23] with the *umap-learn* python package. For Neural Network features extraction, segmented and processed radiograms were used. Based on the classification model architecture, a model which took into consideration all the data available in the database was trained. From the trained model the pre-last layer containing a vector of 261 neural network features was extracted. Beforehand given features were preprocessed by removing features with zero variance and data scaling. Initial dimensionality reduction was performed using Principal Components Analysis (PCA) leaving components that explain 90% of the variance. From acquired features space a UMAP embedding was created using the cosine distance metric.

### Class subtype prediction

Two-dimensional Gaussian mixture model (GMM)[24] was fitted to the 2D UMAP representation of the complete dataset. Initial conditions were set randomly, and the fitting procedure was repeated 100 times. The best model was found using the Bayesian information criterion. To avoid overfitting, the regularization parameter was added with a value of 0.1. Constructed GMM model was also used to predict the class subtype of new unseen data.

### Image super-resolution model

The ESRGAN model[25] was utilized as a base architecture for increasing the resolution of the low-resolution radiographs. The model was trained in a GAN framework, where two neural networks aim to minimize a loss function mimicking a zero-sum game. The loss function was the same as in the original paper. The model received the low-resolution representation of the radiograph and upsampled the image in a patch-wise manner. The input image was divided into patches of 50x50 resolution. The patches were fed into the neural network and the resulting patches were formed into a high-resolution representation. The model was constructed from a series of blocks called Residual Dense Blocks (RDBs) designed to enhance the information flow between layers. The blocks contain residual connections between different abstraction layers, which facilitate the aggregation of global and local visual information.

## Lung segmentation model

The U-Net architecture was chosen for lung segmentation[26]. U-Net is divided into contracting and expansive paths. The contracting track is composed of five blocks of two unpadded convolutional layers (3x3 pixels kernel). All but the last block in the contracting path contains a max-pooling operation (2x2 pixels kernel, a stride of 2) to downsample the input image. The expansive path consists of four blocks of a transposed convolutional layer (2x2 pixel kernel, a stride of 2), two convolutional layers (3x3 pixels kernel), and a corresponding feature map from the contracting path. Scaled Exponential Linear Unit (SELU) was applied as an activation function to each but the last convolutional layer of the network which was followed by a sigmoid activation function to generate a segmentation map for the input image. A 10% dropout[27] was introduced after each max-pooling layer in the model to prevent the model from overfitting and to cover for model uncertainty. Additionally, batch normalization[28] was applied after each but last convolution layer. Data augmentation was used to increase the final model generalization. Images from the training set were flipped horizontally, rotated by 0.1 radians in both directions, and zoomed by a factor of 0.1.

During network training, resized and padded input images and their corresponding segmentation maps were provided to train the network. Sorensen-Dice coefficient (SDC) served as a similarity measure and was used in loss function minimized during the training. For an optimization algorithm of the network, a first-order gradient-based method called ADAM (adaptive learning rate optimization algorithm) was chosen as being robust for many non-convex optimization problems. Xavier uniform initialization was selected for the network's parameter initialization[29].

During the post-processing phase, the segmentation map produced by U-Net was further converted to ensure whole lung extraction and rejection of any non-lung regions. This phase can be decomposed into several steps. First, the system maintained the two largest segmentation regions and discarded any additional separate segmented regions. Next, the morphological operation of the opening was applied to smoothen the contours of the lungs. The morphological operation of closing was applied to fill in any missing pixels inside the segmented region. Finally, the output image was a product of the post-processed mask and the input CXR. The generated lung segmentation undergoes further processing to remove any artifacts and improve its quality by adding a convex hull. After applying the lung mask, all pixels not belonging to the lung area were removed from the original image. The lung pixels were standardized but the thresholds for outliers have been lowered to 0.0005 and 0.9995 percentiles. The next step was a ROI modification - both lungs were identified in the image and then placed at a minimum distance from each other. At the same time, the image was cropped on each side to the lung border to reduce the background participation in the classification process.

Every image was resized to 512x512 pixels while maintaining the aspect ratio to retain the shape of the chest. In cases of different dimensions, padding was added. Standardization was performed by mean subtracting per every pixel and scaling to unit variance. Mean and standard deviation have been saved as pre-processing parameters to perform analogous standardization to new, unseen data, before feeding them to the network.

### Image-based classification model

The output lung image was fed into the next network to perform a diagnosis of normal, pneumonia, or COVID-19 class. The InceptionV3 convolution network was constructed and implemented[30]. This network consists of 17 convolution layers and one fully connected layer. The convolution layers were grouped into four blocks of 4 convolution layers - identical in structure. To accelerate network weight training, neuron weights were initialized using a pre-trained model on the ImageNet dataset. The parameters of the network model (such as batch size, and several epochs) were selected in the process of cross-validation by a factor of three. Finally, the deep learning model was trained in 100 epochs, using a 5-fold cross-validation scheme (5x100 epochs). ADAM algorithm was chosen again to solve the optimization problem in the network learning process (backpropagation). The dataset was artificially expanded using image transformations (rotation, pixel shift, reflection) before their use in each epoch. The CIRCA decision was based on the maximum probability rule.

### Radiomics-based classification model

For radiomic features[31] extractions, an original radiogram, and its lung segmentation were used. Patient lungs were analyzed in three segments, namely LL (Lower Lung), ML (Middle Lung), and UL (Upper Lung). To obtain the mentioned segments, lung segmentation was split evenly into three parts. From each segment, radiomic features were calculated using pyradiomics python package[32].

For each image, a set of 261 radiomic features from the three lung segments was computed. Kruskal-Wallis's test was performed to identify features significant for the differentiation of three classes (normal, pneumonia, and COVID-19), and eta square was used as a measure of the effect size. Examination of the effect size distribution and the impact of chosen subsets of features on the results reduced the number of features to 200 (all features with at least a small effect size). The subset was used to train a dense neural network to differentiate between patient categories. To enlarge the dataset, data augmentation was used – the same set of features was calculated but with the bin width parameter set to 0.01.

The neural network was built of seven dense layers containing 1024, 512, 256, 128, 64, 32, and 3 neurons. All but the last layer had a ReLU activation function, and Nadam was the optimizer. The last layer activation was Softmax. After each layer, a dropout of 20% of randomly selected neurons was applied. The L2 regularization penalty was applied to the kernels. During training, a learning rate of 0.001 and a batch size of 128 were used. The optimal set of parameters was found through hyperparameter tuning. To balance the dataset and to emphasize the COVID-19 cases, the class weights were set as follows: normal 0.1, pneumonia 0.3, and COVID-19 0.9. Features were standardized to have a zero mean and a standard deviation of one.

### Aggregation of multiple models' predictions

The last step of the classification system is the integration of predictions from two CXR classification models. For each image, a vector containing predicted probabilities for each class from the image network and radiomics network was created. Different approaches were considered for the prediction aggregation. The best model, which was a Decision Tree, as well as the optimal parameters were found through hyperparameter tuning. The Decision Tree had a maximum depth of 7, and the quality of a split was measured with the Gini impurity criterion. For each split, the model used three of the features. Each leaf node had a minimum of 100 samples. During the training, class weights were used: normal 0.1, pneumonia 0.3, and COVID-19 0.9.

### CIRCA portal Implementation

The CIRCA system was implemented and made publicly available at https://covid.aei.polsl.pl. The unregistered user can upload CXR and obtain the risk estimates for normal, other pneumonia, and COVID-19 diagnoses. The segmented lung and UMAP with disease subtype prediction are also provided. For the registered users, the option for uploading the verified cases in both, jpg and DICOM standards is available.

### Results

To construct the reliable and complete classification system, we used three different datasets (clinical POLCOVID database, a collection of images found mostly online called COVIDx set, and clinical AIforCovid database) and developed the multi-step procedure (**Figure 1**). First, in a process of data cleaning, we identified and removed low-quality images and duplicates. Then, we constructed a model to identify a clinically relevant region of the CXR that contains the lungs (**Figure 1a**). Next, we divided each class (normal, pneumonia, and COVID-19) into sub-classes using an unsupervised approach, which allowed us to construct a data division scheme. The full classification pipeline (**Figure 1b)** was evaluated during cross-validation and on hold-out sets. At last, the final classification model was tested on an independent test set.

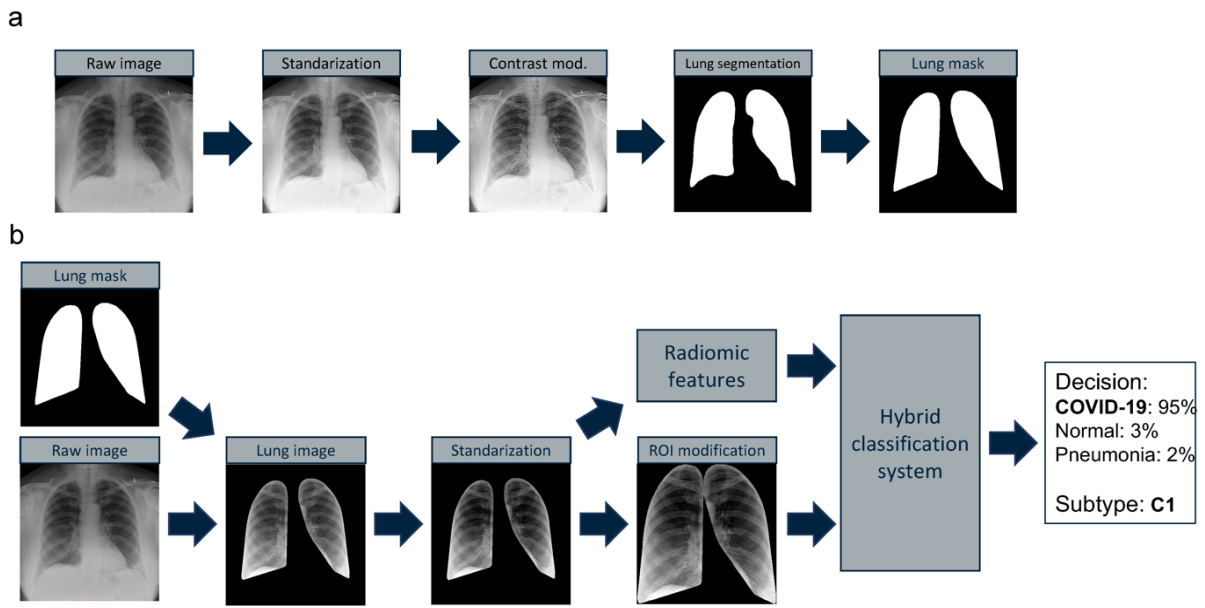

**Figure 1. Description of the subsequent steps of the CIRCA algorithm.** (a) Lung region segmentation includes image pre-processing and deep learning-based segmentation. As a result, a lung mask is obtained. (b) Image classification procedure includes region-of-interest (ROI) definition, calculation of radiomic features, and classification using two models: radiomic-based and image-based. As a result, class prediction with class probability is given along with the estimated class subtype.

## Lung region segmentation for ROI definition

The raw dataset constructed from 3 databases contained 21 317 CXR images (including images of 10 492 normal lungs, 6 720 pneumonia patients, and 4 105 COVID-19 patients). Most of the data were from COVIDx dataset (n=15 403), then from POLCOVID (n=4 809) and from AIforCovid (n=1 105) databases. In COVIDx and POLCOVID images from all three classes were provided in different proportions (COVIDx: 52% normal, 36% pneumonia, 12% COVID-19; POLCOVID: 50% normal, 24% pneumonia, 26% COVID-19), while in AIforCovid only CXRs from COVID-19 patients were given (**Supplementary Table 1**).

The CXR images contain the lung region, together with the heart, blood vessels, airways, and the bones of the chest and spine. However, only the lung region is important for COVID-19 detection, thus we first introduced the lung region segmentation model. For this, we divided the model development data with manually extracted lung masks into three subsets: a training dataset of 1 844 images, a validation dataset of 200 images, and a test dataset of 250 images. CXRs from healthy donors and patients with various types of pulmonary diseases were distributed across training, validation, and test datasets to enhance generalization. The constructed and optimized U-Net network for lung segmentation demonstrated very high accuracy as SDC was equal to 94.86% in the validation dataset, and 93.36% in the testing dataset.

### Inconsistent quality of CXR images

To improve the quality of the data, we first applied our super-resolution model to increase the resolution of all images lower than 512x512 (n=1 694). Only 41% (n=689) of low-resolution data were properly corrected, showing higher similarity to normal resolution images on UMAP-based data visualization (**Supplementary Figure 1a**). We noticed that most of the images that were not significantly improved by the super-resolution model had original resolutions lower than 300x300 pixels, come from the COVIDx database, and/or are from the COVID-19 class. Image upsampling using the proposed model increased the image resolution and quality without introducing any artifacts (**Supplementary Figure 1b**)

Next, we removed 159 images with decreased quality of segmented lung images (segmentation quality score lower than 0.7259). Most low-quality images were from the pneumonia class and COVIDx dataset. We have found various problems with segmented images including poor representation of a single lung or completely unsuccessful segmentation (**Supplementary Figure 2**).

Finally, 20 153 images remained after data cleaning (including images from 10 474 normal lungs, 6 570 pneumonia patients, and 3 109 COVID-19 patients). Again, most of the data were from COVIDx dataset (n=14 261), then from POLCOVID (n=4 793) and from AIforCovid (n=1 099) databases (**Supplementary Table 2**).

## Heterogeneity of lung lesions

Using a clean dataset of 20 153 images, we constructed a 2D UMAP plot on the features from the deep learning model to assess the homogeneity of the data (**Figure 2a**). The density centers of the three classes are distant from each other in the plot, but they are not separated. In each class, some images are located in the region of the other class (**Supplementary Figure 3**). It is assumed, that these images will get lower classification scores than others. We used the 2D GMM model to divide data in each class into 3 subtypes (**Figure 2b**). C1, P1, and N1 are the most different from the others and should contain lung changes (medical features) typical to each class, like consolidations and ground-glass opacities in COVID-19, or no changes in a normal class. Other subtypes might consist of mild symptoms of the disease (**Supplementary Figure 4**).

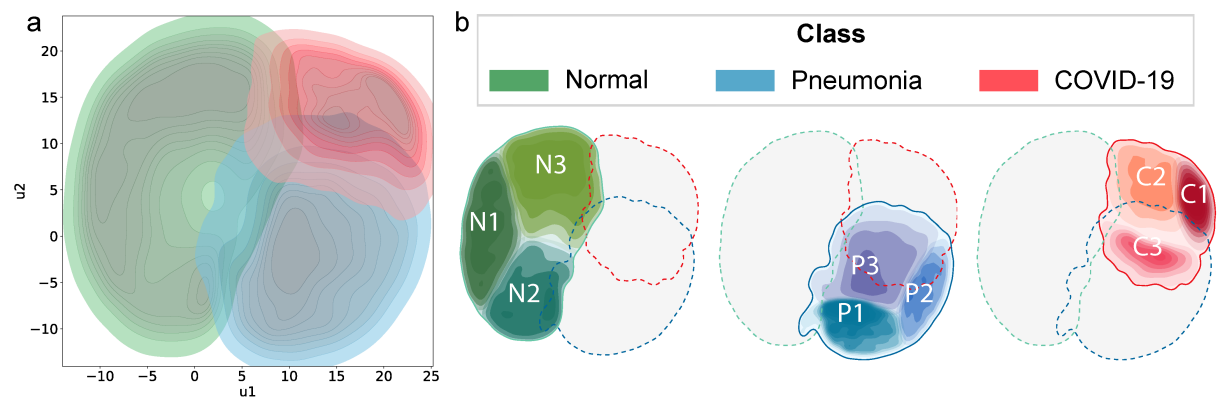

**Figure 2. CXR data heterogeneity.** (a) Density plot in UMAP coordinate space created by including images from 3 color-coded classes (Normal, Pneumonia, COVID-19) and 3 datasets: POLCOVID, COVIDx, and AIforCovid. (b) Results of grouping CXRs into 3 subtypes per class. NX represents subtype X from the Normal class, PX from the pneumonia class, and CX from the COVID-19 class.

## Development of a comprehensive classification system

We analyzed the frequency of image occurrence in subtypes of each class and found significant differences (**Supplementary Table 3**), which may weigh heavily on the estimates of classification model performance indicators. Thus, we first randomly selected hold-out test cases keeping a similar ratio between classes, and then balanced the training sets using two other datasets (**Figure 3**).

In the process of selecting images for the hold-out test set, we assumed 50 cases per class subtype and dataset. However, in some cases, there were not enough images in our database (e.g., P1 subtype in the POLCOVID dataset). In this situation, we increased the number of images in other subtypes from the same class, to have 150 test images per class and dataset (**Supplementary Table 4**). Finally, the hold-out test set consisted of 1 050 CXRs. To provide similar data distributions in test and train sets, we used Gaussian density functions from the 2D GMM model representing each subtype during the sampling process. Additionally, the model was also verified using an independent test set (termed BIMCV).

The remaining data were used for classification model training. To balance the train data, images from 2 other datasets were added: (i) CXRs of pneumonia patients from the NIH dataset; (ii) artificially created CXRs of COVID-19 patients using a GAN-based model. We assumed that CXRs from normal patients were the easiest to classify, so this class was not enriched with new data. As before, the selection of cases that were added was controlled using the Gaussian density functions from the 2D GMM model and mostly under-represented subtypes were extended. Finally, 2 061 images were added to the pneumonia class from the NIH database controlling the age of patients, and 2 635 images to the COVID-19 class using the GAN-based model (**Supplementary Table 5**). Visual inspection of UMAP plots proves the similarity of added images to the proper subtype and class (**Supplementary Figure 5**). We can also notice, that artificially created images look similar to real data (**Supplementary Figure 6**).

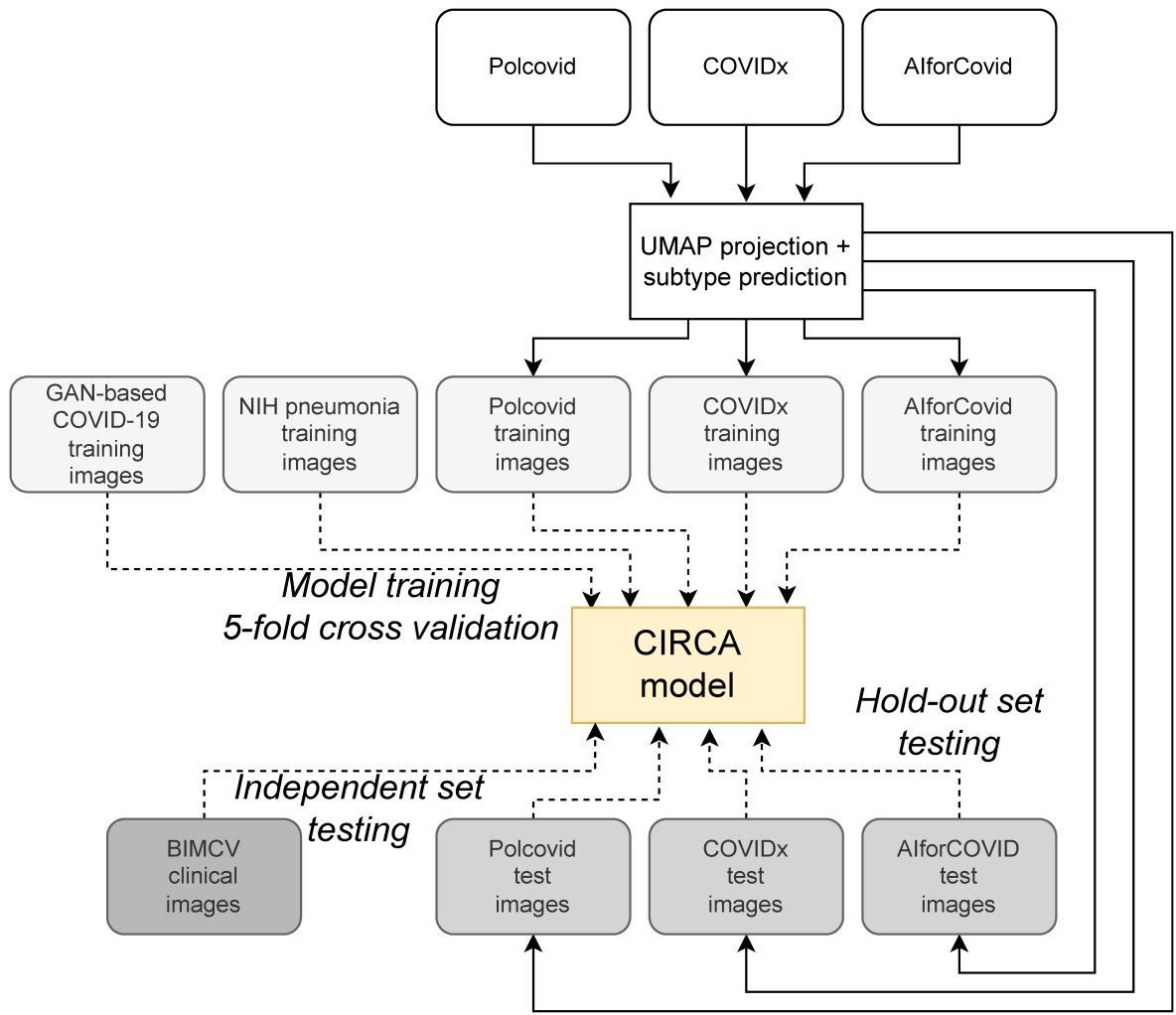

**Figure 3.** Scheme of data division into training and testing sets.

## Evaluation of CIRCA algorithm performance

The performance of the final model was evaluated in three ways: (i) using 5-fold cross-validation during model training; (ii) on hold-out sets from 3 datasets used during training (POLCOVID, COVIDx, and AIforCovid); (iii) on the independent test set (BIMCV).

The average results from cross-validation show the good performance of the CIRCA model (**Table 1**). The accuracy for all classes ranges from 91% to 93%, keeping slightly higher specificity than sensitivity. Also, NPV was higher than PPV for each class to guarantee a low rate of false negative cases. Since the model was constructed to prefer COVID-19 detection, the best results were obtained for this class of images.

**Table 1**. CIRCA classification performance in 5-fold cross-validation scheme. For each measure and class, average values with standard error are shown.

| Index | Normal | Pneumonia | COVID-19 |
|---|---|---|---|
| PPV | 91.31% ± 1.27% | 90.41% ± 1.56% | 80.40% ± 3.51% |
| NPV | 92.42% ± 1.01% | 91.77% ± 0.97% | 97.53% ± 0.90% |
| Sensitivity | 89.69% ± 1.58% | 84.13% ± 2.06% | 91.65% ± 3.23% |
| Specificity | 93.61% ± 1.12% | 95.18% ± 0.86% | 93.53% ± 1.55% |
| Accuracy | 91.93% ± 0.52% | 91.31% ± 0.88% | 93.11% ± 0.83% |

For the hold-out and independent test sets, due to the expected different quality of prediction for the identified subtypes, we performed the quality assessment at the subtypes level and the final indicators at the class level were calculated as weighted averages of partial assessments (**Figure 4**).

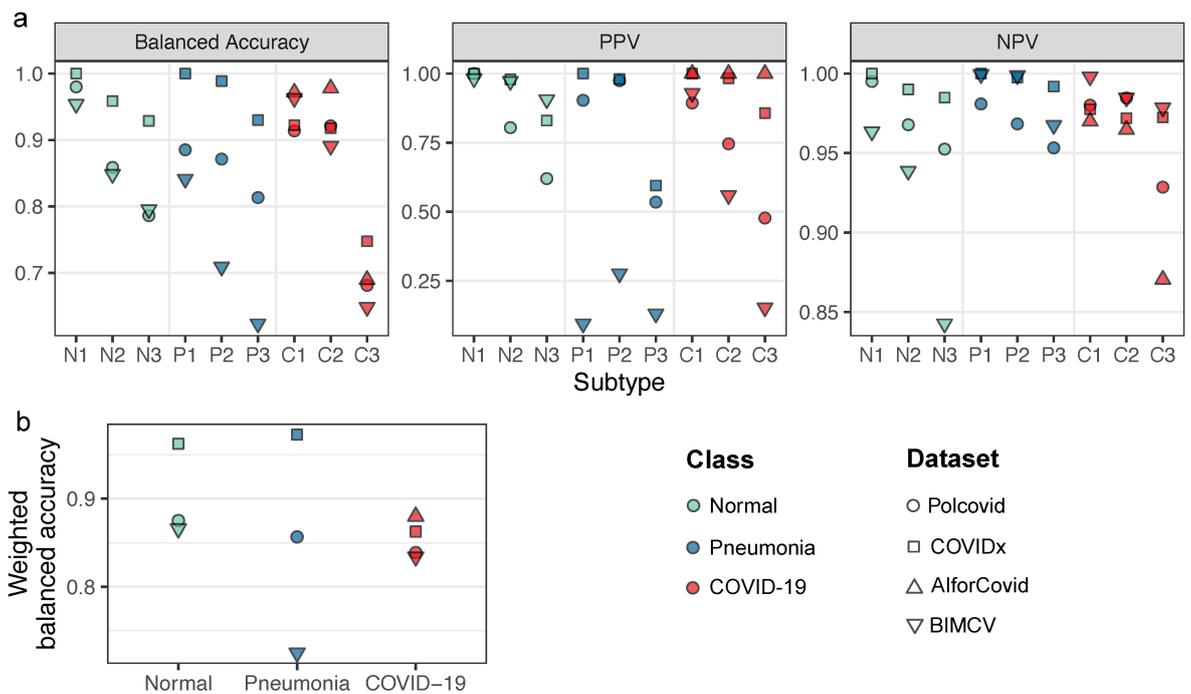

**Figure 4. Evaluation of CIRCA predictive performance in each dataset per subtype (a) and class (b).** Class categories are color-coded, while shapes represent dataset.

The balanced accuracy ranges between 68% and 100% (**Figure 4a**). As in cross-validation, NPV was higher than PPV. The highest performance in terms of accuracy, PPV, and NPV in each class was obtained for the subtypes N1, P1, and C1. Images that were assigned to subtypes N3, P3, or C3 were much harder to be correctly predicted, but in most cases, the predicted subtypes have been confused among themselves (e.g., an image from subtype C3 was classified as P3 or N3). Further, we generated Grad-CAM plots to visualize regions that influenced the classification model decision the most in each dataset and subtype (**Figure 5**). Comparing datasets, higher classification performance was obtained for images from COVIDx than POLCOVID or AIforCovid. The CIRCA model gave a similar performance on the independent BIMCV dataset for normal and COVID-19 class subtypes, but not for pneumonia class subtypes. UMAP-based visualization showed that BIMCV images from pneumonia patients are more similar to COVID-19 or normal cases than in our training data (**Supplementary Figure 7**).

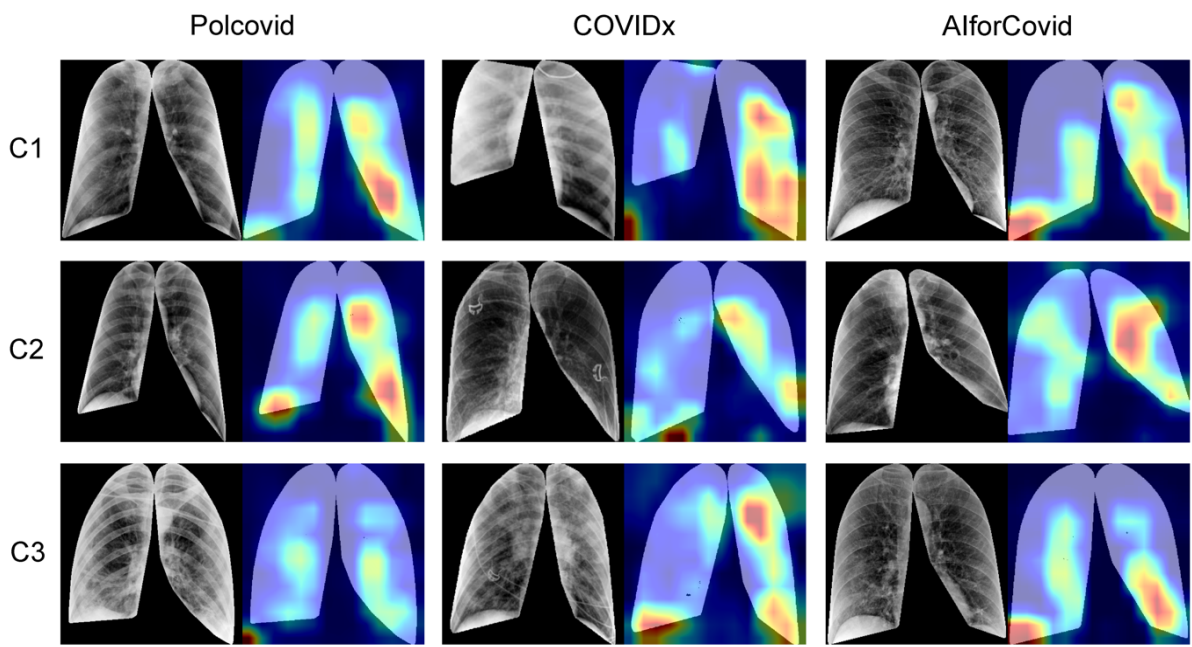

**Figure 5. Confirmation of rational image-based prediction by CIRCA for COVID-19 patients).** For each dataset (in columns) and subtype (in rows) on the left side pre-processed CXRs are shown, while on the right side Grad-CAM plots represent regions that influenced the classification model decision.

The results were also estimated for each class (**Figure 4b** and **Table 2**). For normal and pneumonia classes, images from COVIDx were better classified than images from POLCOVID or BIMCV datasets in terms of all performance measures. For the COVID-19 class, the results were similar between datasets (**Figure 4b**). In all datasets, we obtained high NPV (>85%) and PPV (>70%).

**Table 2.** CIRCA class prediction performance in hold-out test sets.

| Index | Normal | | | Pneumonia | | | COVID-19 | | |
|---|---|---|---|---|---|---|---|---|---|
| | POLCOVID | COVIDx | AIforCovid | POLCOVID | COVIDx | AIforCovid | POLCOVID | COVIDx | AIforCovid |
| PPV | 80.56% | 93.33% | - | 71.79% | 81.56% | - | 71.33% | 97.52% | - |
| NPV | 88.89% | 96.67% | - | 87.07% | 98.52% | - | 85.67% | 90.27% | - |
| Sensitivity | 77.33% | 93.33% | - | 74.67% | 97.33% | - | 71.33% | 78.67% | 84.00% |
| Specificity | 90.67% | 96.67% | - | 85.33% | 89.00% | - | 85.67% | 99.00% | - |

## Model validation in radiologist-based COVID subgroups from an independent dataset

Images from COVID-19 patients in the BIMCV dataset were additionally annotated by expert radiologists based on the visual inspection of CXR into 4 categories: (i) typical appearance; (ii) atypical appearance; (iii) indeterminate appearance; (iv) negative for pneumonia.

We used this annotation to further validate the CIRCA model prediction per class (**Figure 6a**) and subtype (**Figure 6b**). Seventy-six percent of COVID-19 patients wrongly classified as normal cases were in the 'Negative for Pneumonia' category, while correctly classified patients were mostly in the 'Typical Appearance' category. There were 953 images (55%) with no disease symptoms based on radiologist opinion, where CIRCA model gave proper prediction. On the subtype level, most of the cases wrongly classified as normal were in N3, while most of the cases wrongly classified as pneumonia were in P3. From the COVID-19 subtypes, C2 was predicted most often (n=3161).

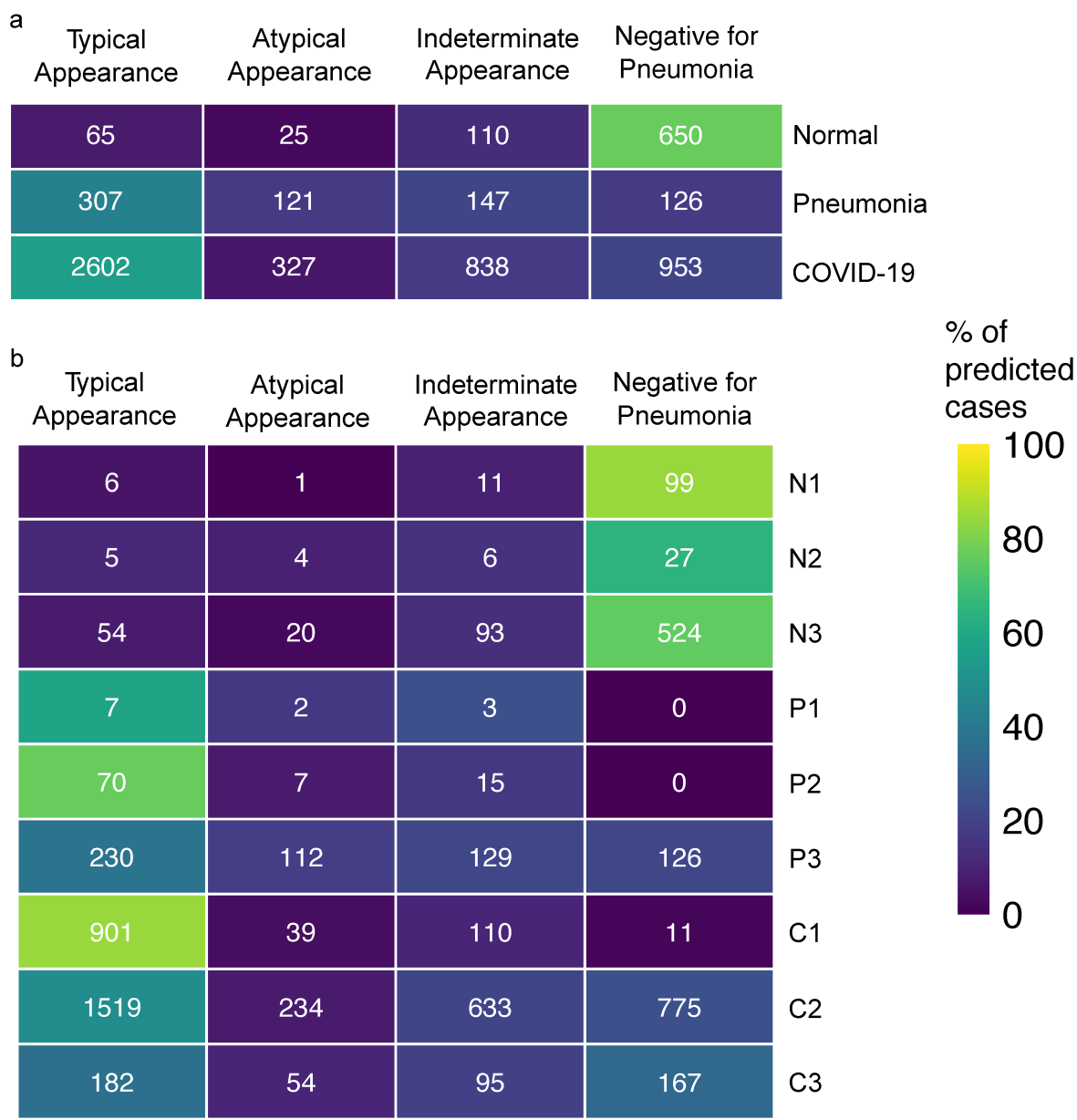

Figure 6. Evaluation of CIRCA subtypes in radiologist-based COVID groups per subtype (a) and class (b). Color coding shows the percent of patients predicted for a given class.

## Comparison to COVID-Net algorithm

Among the models presented in the literature, we searched for an algorithm that fulfills the following assumptions: (i) it is a deep learning-based model; (ii) it is publicly available to download, and it is possible to re-train the model on a custom dataset; (iii) it performs classification into 3 classes including healthy individuals, pneumonia, and COVID-19 patients. COVID-net is the only algorithm that meets all three assumptions. We re-trained the model using our training data and compared the efficiency of CIRCA to COVID-Net using our hold-out test set and an independent test set by calculating differences in different performance metrics on a class level (**Table 3**) and a subtype level (**Supplementary Table 6**). The overall accuracy of CIRCA was around 10 percent higher than COVID-net for all classes. For normal and pneumonia classes, CIRCA gave higher sensitivity and NPV, while for COVID-19 class higher specificity and PPV. F1 index was also better in all cases (**Table 3**). Similar findings were made on the subtype level (**Supplementary Table 6**). Overall, the highest advantage was observed for N3, P3, and C3 subtypes, which consist of class borderline images. Similar results were obtained on the hold-out test set and independent BIMCV dataset, which proves a good generalization of the CIRCA model.

**Table 3.** Gain in class prediction performance after using CIRCA in comparison to COVID-Net in the hold-out test set and an independent test set.

| Dataset | Index | Normal | Pneumonia | COVID-19 |
|---|---|---|---|---|
| TEST | PPV | -4.16% | -9.33% | 24.52% |
| | NPV | 7.28% | 8.99% | -3.37% |
| | Sensitivity | 22.00% | 28.00% | -10.89% |
| | Specificity | -3.07% | -7.47% | 30.00% |
| | F1 | 11.55% | 17.63% | 3.65% |
| | Balanced.Acc | 9.47% | 10.27% | 9.56% |
| BIMCV | PPV | -1.03% | -7.29% | 25.20% |
| | NPV | 14.07% | 1.07% | -0.19% |

| | Sensitivity | 24.96% | 30.67% | -11.35% |
| --- | --- | --- | --- | --- |
| | Specificity | -7.58% | -11.07% | 37.36% |
| | F1 | 18.07% | 9.73% | 23.04% |
| | Balanced.Acc | 8.69% | 9.80% | 13.00% |

## Discussion

The main findings on medical images in patients with COVID-19 include ground-glass opacities, crazy-paving patterns, and areas of consolidation, which usually have bilateral and peripheral distribution and involve multiple lobes. Since the same patterns are often encountered in the pneumonia of other etiologies, including influenza pneumonia, SARS, and MERS pneumonia (as well as respiratory syncytial virus pneumonia, cytomegalovirus pneumonia, adenovirus pneumonia, atypical bacterial pneumonia – mycoplasma pneumonia, chlamydia pneumonia, pulmonary lung edema, and interstitial lung disease), the ability of radiologists to distinguish these diseases is limited, resulting in very low sensitivity of CT (25%)[2,4]. The combination of two deep learning networks working in the sequential mode and a radiomics-based classifier leads to the achievement of effective lung segmentation and disease classification on X-rays, independently of the image quality. The developed CIRCA model worked well during training (5-fold cross-validation), on hold-out sets, and on the independent test set. Even though the classification accuracy was the highest for the COVIDx dataset, as in other solutions proposed in the literature, results on other clinical datasets were also very promising.

There are multiple different deep learning models developed to predict COVID-19 based on CXR images [10,11,17]. Most existing solutions are using pre-trained, supervised learning architectures. While transfer learning gives higher accuracy and faster training in most situations, using models trained on natural images is not the best solution for the analysis of medical images. Further, there are only some solutions that could be run on independent data or could be re-trained on a new dataset. COVID-net algorithm resulted in the highest number of CXRs predicted as COVID-19 disease in comparison to CIRCA, both in the hold-out test set and independent test. Even though such model behavior slightly increased model sensitivity to COVID-19, the resulting NPV was similar, while PPV was much better, showing better usage of CIRCA as a diagnostic test.

Since the SARS-CoV-2 virus appeared very recently, the number of samples available to properly train the model was small in comparison to other classes (healthy, pneumonia). Also, other biases induced by incorrect use of existing datasets or not considering other confounders prevent the translation of prediction models into clinical practice[6,13]. Based on these findings, we gathered two clinically relevant datasets from different regions of the world, in addition to other real data and artificially generated CXR of COVID-19 patients to construct a large representative and well-balanced dataset for model training. Looking at the results of the independent BIMCV dataset, this solution increases the generalization of CIRCA model. Our dataset is publicly available and could be used by other researchers to train or validate upcoming models.

We should stress that in the problem of image-based COVID-19 detection, lung extraction is an essential step[12]. The first deep learning models for COVID-19 diagnosis made their prediction statement based on image artifacts outside lung area, lung size, or other confounders resulting from incorrect dataset usage. Thus, we introduced lung segmentation as an important image pre-processing step. Unfortunately, a relatively low number of training data did not allow us to introduce very deep networks performing both image segmentation and classification tasks.

## Conclusions

The proposed CIRCA system developed on a large representative dataset allowed us to successfully distinguish COVID-19 patients from healthy individuals and other pneumonia cases with a class sensitivity not lower than 92% and specificity above 99% for an independent clinically based test set. The obtained results also meet the requirements for medical screening tools by showing high NPV with a satisfactory PPV rate. The high clinical heterogeneity of the CXR data revealed in the class subtype examination explained the variability of the results obtained. Most COVID-19 patients wrongly classified as normal cases were annotated by radiologists as disease-negative CXRs. Finally, we developed and maintain the online service to provide easy and publicly available access to fast diagnosis support tools.

## Acknowledgments

POLCOVID Study Group: Department of Infectious Diseases and Hepatology, as coordinator: Jerzy Jaroszewicz (Medical University of Silesia in Katowice, Specialised Hospital No. 1 in Bytom), Jan Baron, Katarzyna Gruszczynska (Department of Nuclear Medicine and Image Diagnostics, Medical University of Silesia in Katowice), Magdalena Sliwinska, Mateusz Rataj, Przemyslaw Chmielarz (Voivodship Specialist Hospital in Wroclaw), Edyta Szurowska (II Department of Radiology, Medical University of Gdansk), Jerzy Walecki, Samuel Mazur, Piotr Wasilewski (Central Clinical Hospital of the Ministry of Internal Affairs and Administration in Warsaw), Tadeusz Popiela, Justyna Kozub (Collegium Medicum of the Jagiellonian University in Krakow), Grzegorz Przybylski, Anna Kozanecka (Kujawsko-Pomorskie Pulmonology Center in Bydgoszcz), Andrzej Cieszanowski, Agnieszka Oronowicz-Jaskowiak, Bogumil Golebiewski (National Institute of Oncology in Warsaw, Department of Imaging Diagnostics), Complex of Health Care Centres, Mateusz Nowak (Silesian Hospital in Cieszyn),

GeCONiI infrastructure funded by NCBiR project no. POIG.02.03.01-24-099/13. Additionally, AS and WP are holders of a European Union scholarship through the European Social Fund, grant POWR.03.05.00-00-Z305, and JT is the holder of a European Union scholarship through the European Social Fund, grant no. POWR.03.02.00-00-I029.